\documentclass[aps,prd,twocolumn,preprintnumbers,superscriptaddress,dblfloatfix,nofootinbib]{revtex4-1}
\usepackage[utf8]{inputenc}
\usepackage{lmodern}
\usepackage{amsmath,amssymb}
\usepackage{graphicx}
\usepackage{url}
\usepackage{color}
\usepackage{enumitem}
\usepackage{subfigure}
\usepackage[dvipsnames]{xcolor}
\usepackage[colorlinks=true,breaklinks=true]{hyperref}
\hypersetup{allcolors=[rgb]{0.0 0.0 0.6},linkcolor=[rgb]{0.75 0.05 0.05}}
\usepackage{bm}
\usepackage{epsfig}
\usepackage{slashed}
\usepackage{color}
\usepackage{accents}
\usepackage[dvipsnames]{xcolor}
\usepackage[colorlinks=true,breaklinks=true]{hyperref}
\hypersetup{allcolors=[rgb]{0.0 0.0 0.6},linkcolor=[rgb]{0.75 0.05 0.05}}

\renewcommand\[{\left[}

\newcommand{\exclude}[1]{}

\begin{document}
\preprint{IPMU23-0015}  

\title{Defrosting and Blast Freezing  Dark Matter}
	
\author{Marcos M.  Flores} 
\affiliation{Department of Physics and Astronomy, University of California, Los Angeles \\ Los Angeles, California, 90095-1547, USA}
\author{Chris Kouvaris} 
\affiliation{Physics Division, National Technical University of Athens,\\ 15780 Zografou Campus, Athens,
Greece}
\author{Alexander Kusenko} 
\affiliation{Department of Physics and Astronomy, University of California, Los Angeles \\ Los Angeles, California, 90095-1547, USA}
\affiliation{Kavli Institute for the Physics and Mathematics of the Universe (WPI), UTIAS \\The University of Tokyo, Kashiwa, Chiba 277-8583, Japan}
\affiliation{Theoretical Physics Department, CERN, 1211 Geneva 23, Switzerland}

\date{\today}
	
\begin{abstract}
We show that the present-day dark matter abundance can be produced through a novel mechanism that involves a very rapid thermal freeze-out caused by  inhomogeneous heating and successive fast cooling of small fireballs in the early Universe. The fireballs can be produced from energy deposited in small scale structure growth induced by Yukawa interactions in certain particle species. Yukawa interactions are known to cause  growth of halos  even during a radiation dominated era, and the same interactions facilitate cooling and collapse of the halos by the emission of scalars. Energy deposited in the Standard Model plasma at the locations of the halo collapse can heat the  plasma, re-establishing thermal equilibrium.  The subsequent expansion and cooling of plasma fireballs leads to freeze-out of dark matter on timescales much shorter than the Hubble time. This mechanism can produce the right abundance of dark matter for masses and annihilation cross sections previously thought to be ruled out.

\end{abstract}

\maketitle
	

The weakly interacting massive particle (WIMP) is a well-motivated dark matter (DM) candidate. The most commonly assumed production scenario is based on freeze-out: the DM abundance is frozen at  the temperature at which the WIMP annihilation rate becomes slower than the expansion rate of the Universe.  Thus, it is the Hubble rate that determines the WIMP abundance. 

However, there may be another relevant timescale affecting freeze-out.  A recently discovered phenomenon of halo formation in some particle species during the radiation dominated era~\cite{Amendola:2017xhl, Savastano:2019zpr, Flores:2020drq, Domenech:2021uyx, Flores:2022uzt, Domenech:2023afs} can create inhomogeneous heating of plasma, with subsequent cooling of the produced fireballs, which introduces a new timescale, much shorter than the Hubble timescale.  The WIMP freeze-out is then determined by this shorter timescale, rather than the Hubble rate, leading to a different dependence of the DM abundance on the annihilation cross sections. In this paper we will explore the implications of defrosting and blast-freezing plasma for WIMP abundance.  We will show that this possibility opens a new range of WIMP parameters, which has important implications for direct and indirect DM searches. 

The traditional DM formation scenario involves a heavy particle $X$ which is weakly coupled to the Standard Model (SM) early in the evolution of the Universe. At high temperatures, the $X$ population is initially in thermal equilibrium with the SM. As the Universe expands, the DM abundance is diluted until $XX\leftrightarrow{\rm SM}$ interactions occur slowly compared to the Hubble rate. Once interactions become rare, the comoving  number density of $X$ particles remains fixed to the present day. This ``freeze-out" process is described by the Boltzmann equation,
\begin{equation}
\dot{n}_X + 3H n_X
=
-
\langle
\sigma_{\rm ann} v
\rangle
\left(
n_X^2 - (n_X^{\rm eq})^2
\right),
\end{equation}
where $\langle \sigma_{\rm ann} v \rangle$ is the thermally averaged cross section times the relative particle velocity.
The temperature at which the final DM abundance is frozen out, $T_{\rm FO}^{X}$, can be approximated by solving 
\begin{equation}
\Gamma_{\rm ann}(T_{\rm FO}^{X}) 
\equiv 
\langle
\sigma_{\rm ann} v 
\rangle
n_{X}^{\rm eq}(T_{\rm FO}^{X})
=
H(T_{\rm FO}^{X}).
\end{equation}
 The present-day $X$ abundance is given by~\cite{Baumann:2022mni}
\begin{equation}
\label{eq:WIMPAbundance}
\Omega_X \simeq 
5.2\times 10^{-2}
\frac{x_{\rm FO}^{X}}{\sqrt{g_*(M_X)}}
\left(
\frac{10^{-8}\ {\rm GeV}^{-2}}
{\langle\sigma_{\rm ann} v\rangle}
\right),
\end{equation}
where $x_{\rm FO}^X = m_X/T_{\rm FO}^X$. The above result is insensitive to the mass of the heavy particle, and primarily determined by the cross section. The fact that the observed DM particle density is reproduced if
\begin{equation}
\sqrt{{\langle\sigma_{\rm ann} v\rangle}}
\sim
0.1\sqrt{G_F},
\end{equation}
where $G_F$ is the Fermi four-interaction strength, is referred to as the \textit{WIMP miracle}.

Simultaneously, Eq. \eqref{eq:WIMPAbundance} illustrates the fact that larger cross sections can lead to an underabundance of DM. The principle goal of this paper is to introduce a scenario which can produce the proper DM abundance, particularly in the high-annihilation cross section parameter space  where the traditional freeze-out formalism produces too little DM, thus failing to generate the abundance observed today.

This scenario relies on the recently explored principle of \textit{early structure formation}. The generation of overdensities by long-range forces, like Yukawa interactions, has been examined in many contexts~\cite{Amendola:2017xhl, Savastano:2019zpr, Flores:2020drq, Domenech:2021uyx, Flores:2022uzt, Domenech:2023afs}. Yukawa forces are generally stronger than gravity, thus allowing for the formation of structure during both the matter and radiation dominated eras. The growth of overdensities through Yukawa forces can lead to bound states~\cite{Wise:2014jva,Gresham:2017cvl,Gresham:2018rqo}, or in the presence of radiative cooling via the same Yukawa interaction, collapse and formation of  primordial black holes~\cite{Flores:2020drq,Flores:2021jas}. Alternatively, an overdensity may evaporate due to annihilation of its constituent particles. However, if these particles couple to the SM, the formation, collapse, and annihilation of an overdense region can locally heat the SM plasma. This inhomogeneous, local heating has proven useful when applied to either the matter antimatter asymmetry of the Universe~\cite{Asaka:2003vt, Flores:2022oef} or the generation of primordial magnetic fields~\cite{Durrer:2022cja}.

\begin{figure}[h!]
\centering
\includegraphics[width=0.4\textwidth]{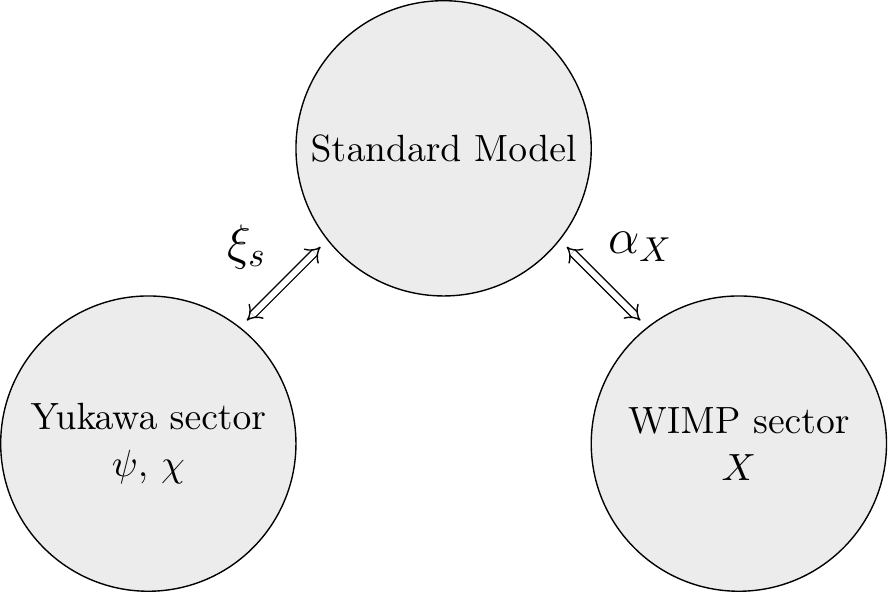}\\
\caption{Schematic diagram of the sectors utilized in our scenario. Here,
$\xi_s$ is the efficiency of energy transfer from the Yukawa sector to the SM [see discussion below Eq.~\eqref{eq:InitialTemp}], and $\alpha_X$ is the WIMP to SM coupling.
}
\label{fig:sector_diag}
\end{figure}

Following Ref. \cite{Flores:2020drq} we consider a dark (sub)sector with a heavy fermion $\psi$ and a light scalar $\chi$ interacting via Yukawa coupling: 
\begin{equation}
\mathcal{L}
\supset 
\frac{1}{2}m_\chi^2\chi^2
+
y\chi\bar{\psi}\psi + \mathcal{L_{\rm Y-SM}}.
\label{eq:Yukawa_sector}
\end{equation}
This sector, which we will refer to as the Yukawa sector throughout, is introduced in addition to the SM and the WIMP $X$ (which may be accompanied by some additional new physics).  The interactions in the Yukawa sector of Eq.~(\ref{eq:Yukawa_sector}) are designed to create the inhomogeneous heating.  The Yukawa sector is weakly coupled to the SM via the cross terms in $ \mathcal{L_{\rm Y-SM}}$ and is not directly linked to the WIMP sector. A schematic of the individual sectors can be found in Fig.~\ref{fig:sector_diag} .  We will parametrize the strength of the SM-to-Yukawa sector coupling below, when we discuss the energy transfer from the Yukawa fireballs to the SM plasma. 

We require that  the fermions $\psi$ are either stable or have a total decay width $\Gamma \ll m_\psi^2/M_{\rm Pl}$ where $M_{\rm Pl}$ is the reduced Planck mass with numerical value, $M_{\rm Pl}
\approx 2.4\times 10^{18}$ GeV.  This ensures there is a cosmological epoch where the $\psi$ particles can become nonrelativistic, decoupled from equilibrium, and interact via the long-range force mediated by the $\chi$ field.

The strength of the Yukawa interaction is generally much larger than gravity. This is demonstrated by comparing the strength of each force, i.e., through $\beta\equiv yM_{\rm Pl}/m_\psi\gg 1$. It should also be briefly noted that another key difference between Yukawa interactions and gravity is the fact that Yukawa interactions couple to the number density of $\psi$ rather  than its energy density.

\begin{figure}[h!]
\centering
\includegraphics[width=0.45\textwidth]{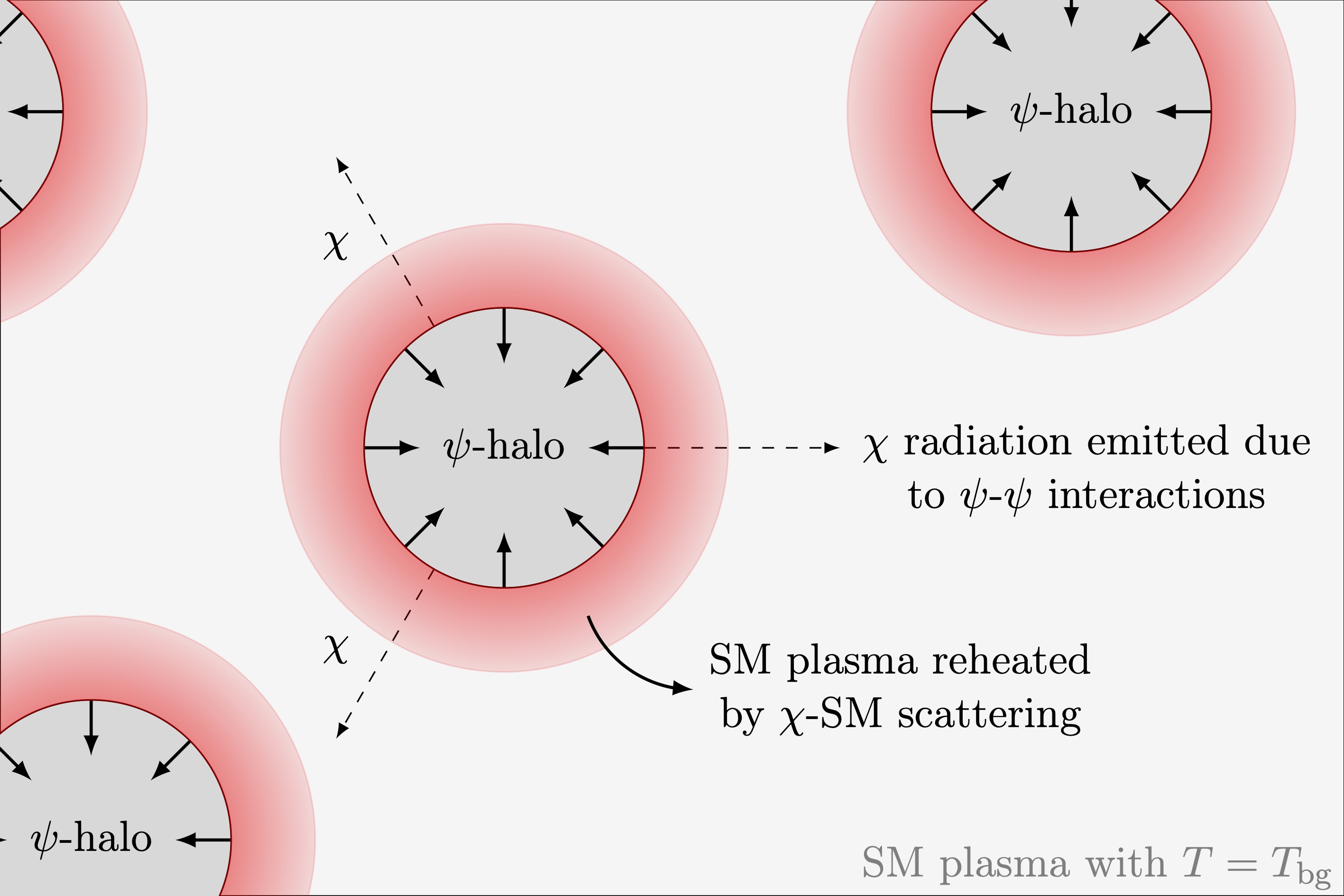}\\
\caption{A visualization of $\psi$ halo collapse which leads to inhomogeneous heating.
}
\label{fig:halo_diag}
\end{figure}

The presence of additional long-range scalar forces generically leads to the rapid development of structure as the overdensities $\Delta\equiv (n_\psi-\bar{n}_\psi)/\bar{n}_\psi$ grow considerably fast. In particular, it has be demonstrated that~\cite{Amendola:2017xhl, Savastano:2019zpr, Flores:2020drq, Domenech:2021uyx, Flores:2022uzt}
\begin{equation}
\Delta \propto a^{\beta},
\quad \beta\gg 1,
\end{equation}
where $a$ is the scale factor, even during radiation domination.
For reference, matter perturbations under the influence of gravity grow as $\delta \propto\ln a$ during radiation domination and $\delta \propto a$ during matter domination. The rapid growth of structures is generally faster than the Hubble rate, implying that the overdensities become nonlinear within a Hubble time. This is followed by the formation of virializied $\psi$ halos comprised of $\psi$ particles.

The details of this early structure formation have been explored both analytically and numerically in Ref.~\cite{Domenech:2023afs}. In this study, the  quartic terms in the $\chi$ potential are included. Both analytical results and $N$-body simulations point to the possibility of a rapid structure formation in the presence of the background dynamics of the scalar field $\chi$.

For the formation of overdensities to occur, we require that $\bar{\psi}\psi\leftrightarrow \chi\chi$ interactions freeze out so that a fixed population of $\psi$ particles can be captured into $\psi$ halos. To do so, we will use a similar framework as described above and define the $\psi$ freeze-out temperature as the solution to the following equation:
\begin{equation}
\label{eq:PsiFreezeOut}
\frac{\Gamma(T_{\rm FO}^{\psi})}{H(T_{\rm FO}^{\psi})} = 1\ 
{\rm for}
\ 
\Gamma(T)
\simeq
\frac{y^4}{4\pi(T^2 + m_\psi^2)}n_{\psi}^{\rm eq}(T)
.
\end{equation}
Once the temperature of the $\psi$ fluid reaches $T_{\rm FO}^{\psi}$, scalar forces can begin to coalesce material into $\psi$ halos, as previously described.

Before calculating this temperature we must note that unlike with gravitational forces, the binding energy of the Yukawa interactions can contribute to the total energy budget in a nontrivial fashion. To accommodate for this possibility, we include an additional energy component into the equation describing the evolution of the Hubble parameter,
\begin{equation}
3M_{\rm Pl}^2 H^2 = \rho_{\rm rad} + \rho_\psi + \rho_y,
\end{equation}
where $\rho_y$ accounts for the energy density of Yukawa potential energy. The length scale of the scalar force $m_\chi^{-1}$ requires that we consider two regimes, namely when $H^{-1} < m_{\chi}^{-1}$ or $H^{-1} > m_{\chi}^{-1}$. When the horizon is smaller than the Compton wavelength of the mediator, the entire Hubble volume is subject to influence from the scalar interaction. Alternatively, when the horizon grows beyond $m_\chi^{-1}$ only subhorizon regions can communicate via the Yukawa force. In this case the number of regions subject to scalar interactions within the horizon is $N_h = (m_\chi/H)^3$. The relationship between the horizon size and the mediator mass gives two expressions for the Yukawa energy density,
\begin{equation}
\rho_y(T)
=
\frac{3y^2}{4\pi m_\psi^2 H^{-3}}
\begin{cases}
M_{\rm hor}^2/H^{-1} & H^{-1} < m_\chi^{-1}\\
N_h M_{\rm hal}^2/m_\chi^{-1} & H^{-1} > m_\chi^{-1}
\end{cases},
\end{equation}
where
\begin{equation}
\begin{pmatrix}
M_{\rm hor}\\
M_{\rm hal}
\end{pmatrix}
=
\frac{4\pi}{3}m_\psi n_{\psi}^{\rm eq}(T)
\begin{pmatrix}
H(T)^{-3}\\
m_\chi^{-3}
\end{pmatrix}.
\end{equation}
Depending on the selection of $\{m_\psi, y, m_\chi\}$ we have three relevant temperatures. The Universe originally begins from a radiation dominated era which eventually transitions to Yukawa domination at $T_{\rm eq}^{\rm RD\to YD}$. After this, the horizon size grows to exceed $m_{\chi}^{-1}$ at $T_{m_\chi = H}$. Later on as the Universe keeps expanding, the number density of the $\psi$ fluid rapidly decreases as the temperature falls below $m_\psi$. This allows for the reestablishment of radiation domination at $T_{\rm eq}^{\rm YD\to RD}$. This leads to an evolution of the Hubble parameter given by
\begin{equation}
H(T)^2
=
\begin{cases}
	\frac{\pi^2}{90}g_*\frac{T^4}{M_{\rm Pl}^2} &  T \lesssim T_{\rm eq}^{\rm YD\to RD}\ \&\ T \gtrsim T_{\rm eq}^{\rm RD\to YD}\\[0.25cm]
	\frac{2\pi^{1/2}}{3M_{\rm Pl}}yn_\psi^{\rm eq}(T) & T_{m_\chi = H} \lesssim T \lesssim T_{\rm eq}^{\rm RD\to YD}\\[0.25cm]
	\frac{4\pi}{9M_{\rm pl}^2}\frac{y^2n_{\psi}^{\rm eq}(T)^2}{m_\chi^2} & T_{\rm eq}^{\rm YD\to RD} \lesssim T \lesssim T_{m_\chi = H}.
	\end{cases}	
\end{equation}
Our choice of parameters might lead to the situation where $T_{\rm eq}^{\rm YD\to RD} > T_{m_\chi = H}$. In this case the evolution of the Hubble parameter instead follows
\begin{equation}
H(T)^2
=
\begin{cases}
	\frac{\pi^2}{90}g_*\frac{T^4}{M_{\rm Pl}^2} &  T \lesssim T_{\rm eq}^{\rm YD\to RD}\ \&\ T \gtrsim T_{\rm eq}^{\rm RD\to YD}\\[0.25cm]
	\frac{2\pi^{1/2}}{3M_{\rm Pl}}yn_\psi^{\rm eq}(T) & T_{\rm eq}^{\rm YD\to RD} \lesssim T \lesssim T_{\rm eq}^{\rm RD\to YD}.
\end{cases}	
\end{equation}
Having determined the Hubble rate in this general framework, we can now determine when the $\bar{\psi}\psi\leftrightarrow \chi\chi$ freeze out using Eq.~\eqref{eq:PsiFreezeOut}.

Before discussing scalar radiation, we will briefly reiterate the thermal history up to this point. The Yukawa sector is lightly coupled to the SM, for example, through high-dimensional operators which result from a higher energy, UV complete theory. At some temperature $T\gtrsim T_{\rm FO}^{\psi}$, the Yukawa sector and the SM decouple. Once $T \lesssim T_{\rm FO}^{\psi}$, early structure formation and collapse can proceed.

The statement that the SM and the Yukawa sector decouple for $T\gtrsim T_{\rm FO}^{\psi}$ can be understood as follows. Firstly, the smallness of the SM to Yukawa sector coupling may lead to early freeze-out of $\bar{\psi}\psi\leftrightarrow$ SM interactions simply by weakness of this interaction strength. However, we only require that $\bar{\psi}\psi\leftrightarrow$ SM interactions are frozen out once $T\sim T_{\rm FO}^{\psi}$. As is usual for freeze-out calculations, $T_{\rm FO}^{\psi} \sim \mathcal{O}(0.1 - 0.01)\ m_\psi$, implying that the equilibrium number density $n_\psi^{\rm eq}(T_{\rm FO}^\psi)$ is within the regime of Boltzmann suppression. This fact, and the smallness of the SM-Yukawa sector coupling imply that these two sectors should decouple at a temperature for $T\gtrsim T_{\rm FO}^{\psi}$.  The exact details of this decoupling process will require an exact specification of the SM-Yukawa coupling, which we will leave to future work.

Without energy dissipation the formation of virialized $\psi$ halos through Yukawa interactions would be the end of the story, with the newly formed halos either remaining stable or evaporating once the constituent particles decay. However, the same Yukawa interaction allows for the emission of scalar radiation much in the same way as electromagnetic interactions between charged particles allow for energy dissipation. Initially the energy is carried away primarily through $\psi$  pair interactions, i.e., through free-free or bremsstrahlung emission. As the halo continues collapsing, it becomes optically thick to the scalar mediator $\chi$, and radiation becomes trapped. This restricts radiation to be emitted only from the surface. Given that surface cooling is the least efficient channel of energy removal, its associated timescale determines whether radiative cooling is rapid enough to facilitate collapse within a Hubble time. The characteristic timescale associated with the energy loss during the surface radiation stage is
\begin{equation}
\tau_{\rm cool}
\equiv 
\frac{E}{|dE/dt|}
\sim R_h \ll H^{-1},
\end{equation}
thus implying that radiative collapse is swift.

For our exploration of DM generation, we assume that the formation and collapse of the $\psi$ halos occur quickly after $\bar{\psi}\psi\leftrightarrow \chi\chi$ freeze out. Specifically, we assume that the formation and collapse occur rapidly so the change in average background temperature outside of the collapsing halos is negligible. Therefore, we will set the background temperature $T_{\rm bg}$ equal to the $\psi$ freeze-out temperature $T_{\rm FO}^{\psi}$. At formation, we will assume that the halos initially have radius $R_h = m_\chi^{-1}$ and masses
\begin{equation}
M_h 
=
\frac{4\pi}{3}m_\psi n_\psi(T_{\rm bg})R_h^3	
.
\end{equation}
In principle, the masses and radii of the halos should be derived from an underlying distribution such as that described in the Press-Schechter formalism. Existing $N$-body simulations have yet to determine the precise nature of the mass distribution of the $\psi$ halos. Our assumption that the halos have a similar composition is motivated by both simplicity and the fact that the strength of Yukawa interactions will facilitate a rich merger history, as demonstrated in Ref.~\cite{Domenech:2023afs}, which form halos of a maximal radius given by the Compton wavelength of the mediator $\chi$.

Without an asymmetry in the $\psi$ population, the halos will annihilate after the initial stage of collapse. Annihilations will begin when the average distance between particles within the halo is less than the Compton wavelength, i.e.,
\begin{equation}
R_{\rm ann} 
\equiv
\frac{1}{m_\psi}
\left(
\frac{3}{4\pi}\frac{M_h}{m_\psi}
\right)^{1/3}
.
\end{equation}
The energy released through scalar quanta during the initial collapse is given by
\begin{equation}
\Delta E_{\rm emis}
=
\frac{y^2 M_h^2}{m_\psi^2 R_{\rm ann}}
\left(
1 - \frac{R_{\rm ann}}{R_h}
\right)
.
\end{equation}
Here, $\Delta E_{\rm emis}$ designates the energy released through scalar radiation alone. Annihilation of the halo will also release energy into the ambient plasma, $\Delta E_{\rm ann} = \epsilon_{\rm ann} M_h$ where $\epsilon_{\rm ann} \leq 1$ parametrizes the efficiency of annihilation. In particular, $\epsilon_{\rm ann}$ encapsulates deviations from perfect annihilation, which may be caused by many sources like, for example, the nonsphericity of the $\psi$ halo.  The total energy emitted through scalar particles is the sum, $\Delta E\equiv \Delta E_{\rm emis} + \Delta E_{\rm ann}$. We will assume that the Yukawa sector, which contains $\psi$ and $\chi$, is weakly coupled to the SM (see Fig.~\ref{fig:sector_diag}). The sudden release of a large amount of energy from collapse and halo annihilation locally heats the SM plasma (see Fig.~\ref{fig:halo_diag}). We will assume that the collapsing halos become relativistic so that the initial temperature of the heated region is
\begin{equation}
\label{eq:InitialTemp}
T_i^4 = \frac{90\xi_s\Delta E}{4\pi^3g_*(T_i)R_i^3},
\end{equation}
where $R_i$ is the initial radius of the heated region, $g_*(T_i)$ are the relativistic degrees of freedom at $T_i$ and, crucially, $\xi_s$ is the efficiency of energy transfer from the Yukawa sector $\chi$ particles to the SM plasma. Determining the quantity $\xi_s$ from first principles is nontrivial. This parameter could be viewed to encapsulate a number of effects, and not just the coupling from the Yukawa sector to the SM, and further investigations will take a model-dependent approach in order to determine a better understanding of the energy transfer between the SM and the Yukawa sector.

Once heated above the background temperature, the excessive energy spreads out via both a shockwave and diffusion. In the first case the expanding shockwave travels through the SM plasma at the speed of sound. Using energy conservation, we determine the characteristic timescale associated with the explosion to be
\begin{equation}
\tau_{\rm exp}
\equiv
\frac{T}{|dT/dt|}
=
\frac{4R_i}{\sqrt{3}}
\left(
1 + \frac{t - t_i}{\sqrt{3}R_i}
\right)
.
\end{equation}
\begin{figure}[h!]
\centering
\includegraphics[width=0.48\textwidth]{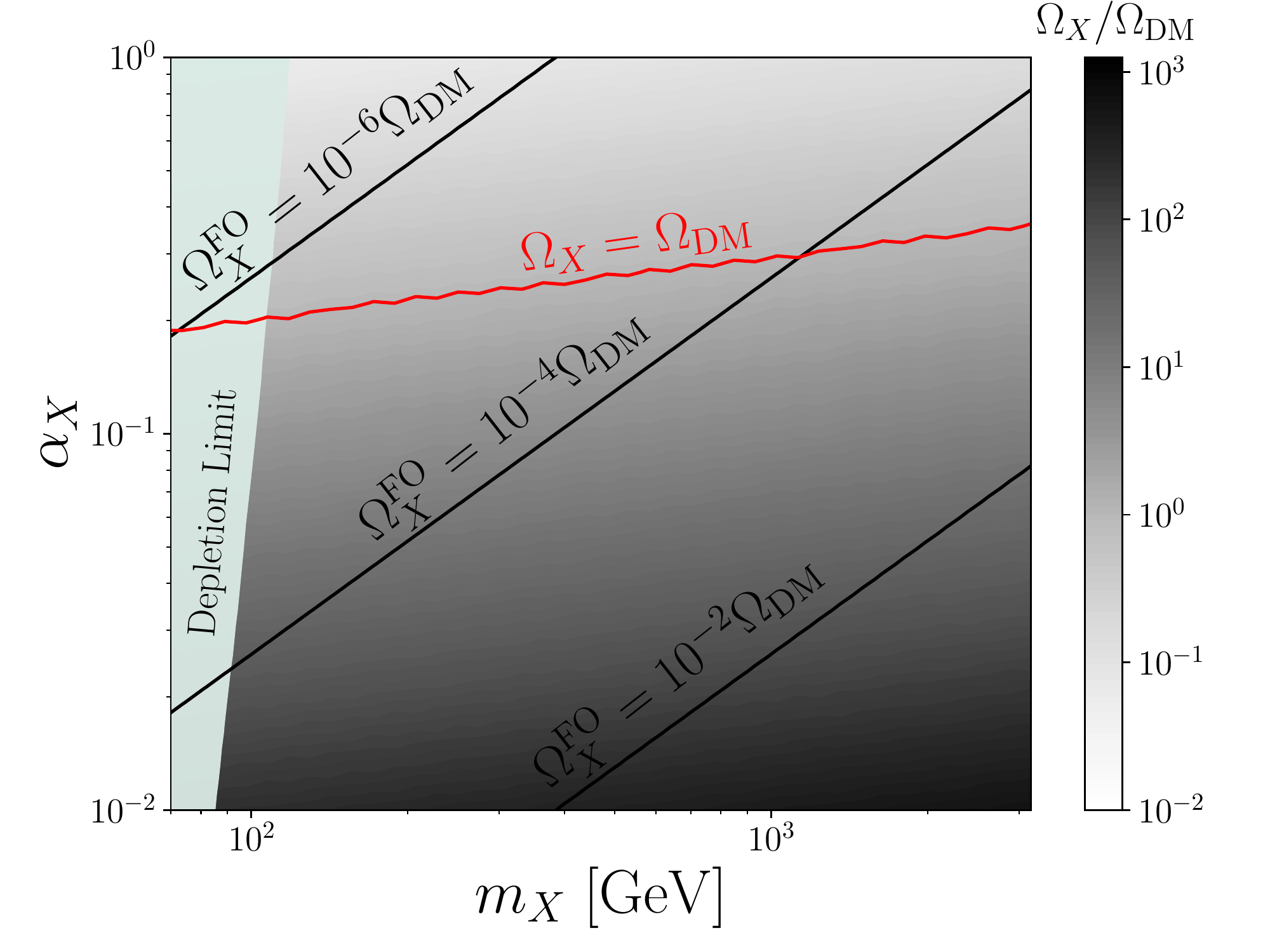}\\
\caption{Predicted DM density for a variety of DM masses and couplings. The red line indicates values within the fireball scenario where the abundance of $X$ is exactly the observed abundance of  DM today. Black contours are the predicted DM density in the conventional freeze-out scenario, i.e., Eq.~\eqref{eq:WIMPAbundance}. The green region is excluded by the depletion limit Eq.~\eqref{eq:DepletionLimit}. Here $y = 0.75$, $m_\psi = 10^{0.3}$ GeV, $m_\chi = 10^{-16}$ GeV, and $\xi_s = 10^{-3}$.}
\label{fig:ParamSpacePlot}
\end{figure}
Most of the energy released during the collapse occurs right before annihilation. Therefore, we approximate $R_i\sim R_{\rm ann}$. The region may also cool through diffusion. The timescale associated with this process is approximately~\cite{Asaka:2003vt}
\begin{equation}
\tau_{\rm diff}
\sim
\frac{R_i^2}{4D}
\left(
\frac{T_i}{T}
\right)^{8/3},
\end{equation}
where $D$ is a diffusion constant. As in Ref.~\cite{Asaka:2003vt} we take $D \sim 1/\gamma_g$ where $\gamma_g\sim 0.3 g_s^2 T$ and $g_s$ is the strong coupling~\cite{Braaten:1990it}. The dissipation timescale is defined as $\tau_{\rm diss} \equiv \min\{\tau_{\rm exp}, \tau_{\rm diff}\}$. Generally, $\tau_{\rm exp} \ll \tau_{\rm diff}$. This means that the expanding fireball is the most rapid method of energy transport. 

Before the early structure formation and collapse, we require that the WIMP sector is decoupled from the SM. Therefore we establish the requirement that the background temperature $T_{\rm bg}\lesssim m_X$. For the parameters we consider the $X$ density produced from traditional freeze-out must also be unable to explain the DM abundance today. 

With these conditions in mind, the collapse of the $\psi$ halos reheats local regions above $T = m_{X}$.  Within these heated regions, thermal equilibrium is reestablished allowing DM production followed by a rapid imhomogeneous refreeze-out. The evolution of the $X$ number density in these heated regions is described by
\begin{equation}
T\frac{d}{dT}
\left(
\frac{n_X(T)}{T^3}
\right)= -\frac{\Gamma_X(T)}{\tau_{\rm diss}^{-1}(T)}
\left[
\frac{n_{X}^2}{T^3} - \frac{(n_{X}^{\rm eq})^2}{T^3}
\right],
\end{equation}
where
\begin{equation}
\Gamma_X(T)
\simeq
\frac{\alpha_X^2}{(m_X^2 + T^2)}n_X^{\rm eq}(T)
.
\end{equation}

Unlike the standard paradigm, here DM freeze-out reoccurs  at the temperature where $\Gamma_X(T_f) = \tau_{\rm diss}^{-1}(T_f)$ where $T_f$ is the freeze-out temperature within an expanding heated region found  by solving the above equation. The fact that the fireballs expand at a much faster rate than the Hubble parameter causes a rapid DM (re-)freeze-out leaving a significant DM abundance even for annihilation cross sections much larger than the weak one required in the standard freeze-out paradigm.  
The resultant DM energy density is thus given by
\begin{equation}
\rho_X(T_{\rm bg}) 
\simeq 
f\cdot m_X n_X^{\rm eq}(T_{f}),
\end{equation}
where $f$ is a volume filling factor defined as
\begin{equation}
f
\equiv
N_h H^3(T_{\rm bg})R_i^3
\left(
\frac{T_i}{T_f}
\right)^{4}.
\end{equation}
We will restrict our parameter space such that $f < 1$, $T_i < M_{\rm Pl}$, and $T_{i} > m_\psi$, $T_{\rm bg}$. Furthermore, to ensure that the DM produced is cold, or nonrelativistic we require that $T_f < m_X$. Lastly, we need to impose a condition for the produced DM. Once the fireball stops expanding, i.e., once its temperature drops to the background value, we have to guarantee that the generated DM density does not annihilate with a rate higher than the Hubble expansion. In such a case we will once again have a depletion of the DM population. This condition reads as
\begin{equation}
\label{eq:DepletionLimit}
\frac{3\sqrt{10}}{\pi}
g_*^{-1/2}
\frac{T_{\rm bg} M_{\rm Pl}}{\tau_{\rm diss}(T_f ) }
 < T_f^3
.
\end{equation}

In Fig.~\ref{fig:ParamSpacePlot} we show the present-day DM abundance as a function of the DM mass, $m_X$, and annihilation coupling, $\alpha_X$. The black contours illustrate the abundance predicted by the traditional freeze-out scenario, where we have taken the thermal averaged cross section to be
\begin{equation}
\langle
\sigma_{\rm ann} v
\rangle
\sim \alpha_X^2/m_X^2.
\end{equation}
The \textit{red line} denotes the region of parameter space within our setup where $\Omega_X = \Omega_{\rm DM}$. The green region indicates parameter space which fails the depletion limit, Eq.~\eqref{eq:DepletionLimit}.  Figure~\ref{fig:ParamSpacePlot} illustrates that our scenario can generate the full abundance of DM in regions of parameter space which are ruled out in the standard freeze-out paradigm. In the presented example, we have used $y = 0.75$, $m_\psi = 10^{0.3}$ GeV, $m_\chi = 10^{-16}$ GeV, and $\xi_s = 10^{-3}$. This slice of parameter space satisfies the conditions discussed previously, but is far from the only acceptable set of parameters.

As touched upon in \cite{Flores:2020drq, Flores:2022oef}, the scalar $\chi$ may act as an additional relativistic degree of freedom thus leading to a modification to $\Delta N_{\rm eff}$. This is permissible, and may even help resolve the Hubble tension~\cite{Bernal:2016gxb,Gelmini:2019deq,Anchordoqui:2019yzc,Vattis:2019efj,Escudero:2019gvw,Gelmini:2020ekg,Vagnozzi:2019ezj,Wong:2019kwg, Perivolaropoulos:2021jda}. Crucially, the collapse of $\psi$ halos is generically asymmetric, thus leading to a nonvanishing quadrupole  moment. This naturally implies the generation of gravitational waves, and studies have shown that such a signal may be detectable by future gravitational wave observatories~\cite{Flores:2022uzt}. 

We note that any inhomogeneities in the production of DM  might occur on such small scales that they have no implications for conventional structure formation or isocurvature fluctuations. The mechanism described above occurs far before matter-radiation equality. In fact, for the parameter space we have considered, the generation of DM occurs before neutrino decoupling. The scales upon which the DM homogeneities are produced are smaller than the horizon size at that time, and are significantly smaller than those which could ultimately impact the formation of large scale structures.

Here we demonstrate that early structure formation from Yukawa interactions can lead to the generation of DM. This occurs due to the collapse of halos of heavy fermions due to scalar radiation, which exchanges energy with SM, thus allowing for local heating of the background SM plasma. These inhomogenous hot spots re-establish thermal equilibrium, allowing for the local formation of DM. Freeze-out, now dependent on the timescale pertaining to the rapid expansion of the heated region, produces a non-negligible abundance of matter which may present itself as the DM observed today. This is particularly applicable to regions of parameter space which are  ruled out in the traditional freeze-out paradigm.

\begin{acknowledgments}
The work of A.K. and M.M.F was supported by the U.S. Department of Energy (DOE) Grant No. DE-SC0009937. 
M.M.F was also supported by the University of California, Office of the President Dissertation Year Fellowship and donors to the UCLA Department of Physics \& Astronomy.  A.K. was also supported by World Premier International Research Center Initiative (WPI), MEXT, Japan, and by Japan Society for the Promotion of Science (JSPS) KAKENHI Grant No. JP20H05853. This work used computational and storage services associated with the Hoffman2 Shared Cluster provided by UCLA Institute for Digital Research and Education’s Research Technology Group.
\end{acknowledgments}


\bibliography{biblio}

 \newcommand{\noop}[1]{}
\begin{thebibliography}{24}%
\makeatletter
\providecommand \@ifxundefined [1]{%
 \@ifx{#1\undefined}
}%
\providecommand \@ifnum [1]{%
 \ifnum #1\expandafter \@firstoftwo
 \else \expandafter \@secondoftwo
 \fi
}%
\providecommand \@ifx [1]{%
 \ifx #1\expandafter \@firstoftwo
 \else \expandafter \@secondoftwo
 \fi
}%
\providecommand \natexlab [1]{#1}%
\providecommand \enquote  [1]{``#1''}%
\providecommand \bibnamefont  [1]{#1}%
\providecommand \bibfnamefont [1]{#1}%
\providecommand \citenamefont [1]{#1}%
\providecommand \href@noop [0]{\@secondoftwo}%
\providecommand \href [0]{\begingroup \@sanitize@url \@href}%
\providecommand \@href[1]{\@@startlink{#1}\@@href}%
\providecommand \@@href[1]{\endgroup#1\@@endlink}%
\providecommand \@sanitize@url [0]{\catcode `\\12\catcode `\$12\catcode
  `\&12\catcode `\#12\catcode `\^12\catcode `\_12\catcode `\%12\relax}%
\providecommand \@@startlink[1]{}%
\providecommand \@@endlink[0]{}%
\providecommand \url  [0]{\begingroup\@sanitize@url \@url }%
\providecommand \@url [1]{\endgroup\@href {#1}{\urlprefix }}%
\providecommand \urlprefix  [0]{URL }%
\providecommand \Eprint [0]{\href }%
\providecommand \doibase [0]{http://dx.doi.org/}%
\providecommand \selectlanguage [0]{\@gobble}%
\providecommand \bibinfo  [0]{\@secondoftwo}%
\providecommand \bibfield  [0]{\@secondoftwo}%
\providecommand \translation [1]{[#1]}%
\providecommand \BibitemOpen [0]{}%
\providecommand \bibitemStop [0]{}%
\providecommand \bibitemNoStop [0]{.\EOS\space}%
\providecommand \EOS [0]{\spacefactor3000\relax}%
\providecommand \BibitemShut  [1]{\csname bibitem#1\endcsname}%
\let\auto@bib@innerbib\@empty
\bibitem [{\citenamefont {Amendola}\ \emph {et~al.}(2018)\citenamefont
  {Amendola}, \citenamefont {Rubio},\ and\ \citenamefont
  {Wetterich}}]{Amendola:2017xhl}%
  \BibitemOpen
  \bibfield  {author} {\bibinfo {author} {\bibfnamefont {L.}~\bibnamefont
  {Amendola}}, \bibinfo {author} {\bibfnamefont {J.}~\bibnamefont {Rubio}}, \
  and\ \bibinfo {author} {\bibfnamefont {C.}~\bibnamefont {Wetterich}},\ }\href
  {\doibase 10.1103/PhysRevD.97.081302} {\bibfield  {journal} {\bibinfo
  {journal} {Phys. Rev. D}\ }\textbf {\bibinfo {volume} {97}},\ \bibinfo
  {pages} {081302} (\bibinfo {year} {2018})},\ \Eprint
  {http://arxiv.org/abs/1711.09915} {arXiv:1711.09915 [astro-ph.CO]}
  \BibitemShut {NoStop}%
\bibitem [{\citenamefont {Savastano}\ \emph {et~al.}(2019)\citenamefont
  {Savastano}, \citenamefont {Amendola}, \citenamefont {Rubio},\ and\
  \citenamefont {Wetterich}}]{Savastano:2019zpr}%
  \BibitemOpen
  \bibfield  {author} {\bibinfo {author} {\bibfnamefont {S.}~\bibnamefont
  {Savastano}}, \bibinfo {author} {\bibfnamefont {L.}~\bibnamefont {Amendola}},
  \bibinfo {author} {\bibfnamefont {J.}~\bibnamefont {Rubio}}, \ and\ \bibinfo
  {author} {\bibfnamefont {C.}~\bibnamefont {Wetterich}},\ }\href {\doibase
  10.1103/PhysRevD.100.083518} {\bibfield  {journal} {\bibinfo  {journal}
  {Phys. Rev. D}\ }\textbf {\bibinfo {volume} {100}},\ \bibinfo {pages}
  {083518} (\bibinfo {year} {2019})},\ \Eprint
  {http://arxiv.org/abs/1906.05300} {arXiv:1906.05300 [astro-ph.CO]}
  \BibitemShut {NoStop}%
\bibitem [{\citenamefont {Flores}\ and\ \citenamefont
  {Kusenko}(2021)}]{Flores:2020drq}%
  \BibitemOpen
  \bibfield  {author} {\bibinfo {author} {\bibfnamefont {M.~M.}\ \bibnamefont
  {Flores}}\ and\ \bibinfo {author} {\bibfnamefont {A.}~\bibnamefont
  {Kusenko}},\ }\href {\doibase 10.1103/PhysRevLett.126.041101} {\bibfield
  {journal} {\bibinfo  {journal} {Phys. Rev. Lett.}\ }\textbf {\bibinfo
  {volume} {126}},\ \bibinfo {pages} {041101} (\bibinfo {year} {2021})},\
  \Eprint {http://arxiv.org/abs/2008.12456} {arXiv:2008.12456 [astro-ph.CO]}
  \BibitemShut {NoStop}%
\bibitem [{\citenamefont {Dom\`enech}\ and\ \citenamefont
  {Sasaki}(2021)}]{Domenech:2021uyx}%
  \BibitemOpen
  \bibfield  {author} {\bibinfo {author} {\bibfnamefont {G.}~\bibnamefont
  {Dom\`enech}}\ and\ \bibinfo {author} {\bibfnamefont {M.}~\bibnamefont
  {Sasaki}},\ }\href {\doibase 10.1088/1475-7516/2021/06/030} {\bibfield
  {journal} {\bibinfo  {journal} {JCAP}\ }\textbf {\bibinfo {volume} {06}},\
  \bibinfo {pages} {030} (\bibinfo {year} {2021})},\ \Eprint
  {http://arxiv.org/abs/2104.05271} {arXiv:2104.05271 [hep-th]} \BibitemShut
  {NoStop}%
\bibitem [{\citenamefont {Flores}\ \emph
  {et~al.}(2022{\natexlab{a}})\citenamefont {Flores}, \citenamefont {Kusenko},\
  and\ \citenamefont {Sasaki}}]{Flores:2022uzt}%
  \BibitemOpen
  \bibfield  {author} {\bibinfo {author} {\bibfnamefont {M.~M.}\ \bibnamefont
  {Flores}}, \bibinfo {author} {\bibfnamefont {A.}~\bibnamefont {Kusenko}}, \
  and\ \bibinfo {author} {\bibfnamefont {M.}~\bibnamefont {Sasaki}},\
  }\href@noop {} {\  (\bibinfo {year} {2022}{\natexlab{a}})},\ \Eprint
  {http://arxiv.org/abs/2209.04970} {arXiv:2209.04970 [astro-ph.CO]}
  \BibitemShut {NoStop}%
\bibitem [{\citenamefont {Dom\`enech}\ \emph {et~al.}(2023)\citenamefont
  {Dom\`enech}, \citenamefont {Inman}, \citenamefont {Kusenko},\ and\
  \citenamefont {Sasaki}}]{Domenech:2023afs}%
  \BibitemOpen
  \bibfield  {author} {\bibinfo {author} {\bibfnamefont {G.}~\bibnamefont
  {Dom\`enech}}, \bibinfo {author} {\bibfnamefont {D.}~\bibnamefont {Inman}},
  \bibinfo {author} {\bibfnamefont {A.}~\bibnamefont {Kusenko}}, \ and\
  \bibinfo {author} {\bibfnamefont {M.}~\bibnamefont {Sasaki}},\ }\href@noop {}
  {\  (\bibinfo {year} {2023})},\ \Eprint {http://arxiv.org/abs/2304.13053}
  {arXiv:2304.13053 [astro-ph.CO]} \BibitemShut {NoStop}%
\bibitem [{\citenamefont {Baumann}(2022)}]{Baumann:2022mni}%
  \BibitemOpen
  \bibfield  {author} {\bibinfo {author} {\bibfnamefont {D.}~\bibnamefont
  {Baumann}},\ }\href {\doibase 10.1017/9781108937092} {\emph {\bibinfo {title}
  {{Cosmology}}}}\ (\bibinfo  {publisher} {Cambridge University Press},\
  \bibinfo {year} {2022})\BibitemShut {NoStop}%
\bibitem [{\citenamefont {Wise}\ and\ \citenamefont
  {Zhang}(2014)}]{Wise:2014jva}%
  \BibitemOpen
  \bibfield  {author} {\bibinfo {author} {\bibfnamefont {M.~B.}\ \bibnamefont
  {Wise}}\ and\ \bibinfo {author} {\bibfnamefont {Y.}~\bibnamefont {Zhang}},\
  }\href {\doibase 10.1103/PhysRevD.90.055030} {\bibfield  {journal} {\bibinfo
  {journal} {Phys. Rev. D}\ }\textbf {\bibinfo {volume} {90}},\ \bibinfo
  {pages} {055030} (\bibinfo {year} {2014})},\ \bibinfo {note} {[Erratum:
  Phys.Rev.D 91, 039907 (2015)]},\ \Eprint {http://arxiv.org/abs/1407.4121}
  {arXiv:1407.4121 [hep-ph]} \BibitemShut {NoStop}%
\bibitem [{\citenamefont {Gresham}\ \emph {et~al.}(2018)\citenamefont
  {Gresham}, \citenamefont {Lou},\ and\ \citenamefont
  {Zurek}}]{Gresham:2017cvl}%
  \BibitemOpen
  \bibfield  {author} {\bibinfo {author} {\bibfnamefont {M.~I.}\ \bibnamefont
  {Gresham}}, \bibinfo {author} {\bibfnamefont {H.~K.}\ \bibnamefont {Lou}}, \
  and\ \bibinfo {author} {\bibfnamefont {K.~M.}\ \bibnamefont {Zurek}},\ }\href
  {\doibase 10.1103/PhysRevD.97.036003} {\bibfield  {journal} {\bibinfo
  {journal} {Phys. Rev. D}\ }\textbf {\bibinfo {volume} {97}},\ \bibinfo
  {pages} {036003} (\bibinfo {year} {2018})},\ \Eprint
  {http://arxiv.org/abs/1707.02316} {arXiv:1707.02316 [hep-ph]} \BibitemShut
  {NoStop}%
\bibitem [{\citenamefont {Gresham}\ and\ \citenamefont
  {Zurek}(2019)}]{Gresham:2018rqo}%
  \BibitemOpen
  \bibfield  {author} {\bibinfo {author} {\bibfnamefont {M.~I.}\ \bibnamefont
  {Gresham}}\ and\ \bibinfo {author} {\bibfnamefont {K.~M.}\ \bibnamefont
  {Zurek}},\ }\href {\doibase 10.1103/PhysRevD.99.083008} {\bibfield  {journal}
  {\bibinfo  {journal} {Phys. Rev. D}\ }\textbf {\bibinfo {volume} {99}},\
  \bibinfo {pages} {083008} (\bibinfo {year} {2019})},\ \Eprint
  {http://arxiv.org/abs/1809.08254} {arXiv:1809.08254 [astro-ph.CO]}
  \BibitemShut {NoStop}%
\bibitem [{\citenamefont {Flores}\ and\ \citenamefont
  {Kusenko}(2023)}]{Flores:2021jas}%
  \BibitemOpen
  \bibfield  {author} {\bibinfo {author} {\bibfnamefont {M.~M.}\ \bibnamefont
  {Flores}}\ and\ \bibinfo {author} {\bibfnamefont {A.}~\bibnamefont
  {Kusenko}},\ }\href {\doibase 10.1088/1475-7516/2023/05/013} {\bibfield
  {journal} {\bibinfo  {journal} {JCAP}\ }\textbf {\bibinfo {volume} {05}},\
  \bibinfo {pages} {013} (\bibinfo {year} {2023})},\ \Eprint
  {http://arxiv.org/abs/2108.08416} {arXiv:2108.08416 [hep-ph]} \BibitemShut
  {NoStop}%
\bibitem [{\citenamefont {Asaka}\ \emph {et~al.}(2004)\citenamefont {Asaka},
  \citenamefont {Grigoriev}, \citenamefont {Kuzmin},\ and\ \citenamefont
  {Shaposhnikov}}]{Asaka:2003vt}%
  \BibitemOpen
  \bibfield  {author} {\bibinfo {author} {\bibfnamefont {T.}~\bibnamefont
  {Asaka}}, \bibinfo {author} {\bibfnamefont {D.}~\bibnamefont {Grigoriev}},
  \bibinfo {author} {\bibfnamefont {V.}~\bibnamefont {Kuzmin}}, \ and\ \bibinfo
  {author} {\bibfnamefont {M.}~\bibnamefont {Shaposhnikov}},\ }\href {\doibase
  10.1103/PhysRevLett.92.101303} {\bibfield  {journal} {\bibinfo  {journal}
  {Phys. Rev. Lett.}\ }\textbf {\bibinfo {volume} {92}},\ \bibinfo {pages}
  {101303} (\bibinfo {year} {2004})},\ \Eprint
  {http://arxiv.org/abs/hep-ph/0310100} {arXiv:hep-ph/0310100} \BibitemShut
  {NoStop}%
\bibitem [{\citenamefont {Flores}\ \emph
  {et~al.}(2022{\natexlab{b}})\citenamefont {Flores}, \citenamefont {Kusenko},
  \citenamefont {Pearce},\ and\ \citenamefont {White}}]{Flores:2022oef}%
  \BibitemOpen
  \bibfield  {author} {\bibinfo {author} {\bibfnamefont {M.~M.}\ \bibnamefont
  {Flores}}, \bibinfo {author} {\bibfnamefont {A.}~\bibnamefont {Kusenko}},
  \bibinfo {author} {\bibfnamefont {L.}~\bibnamefont {Pearce}}, \ and\ \bibinfo
  {author} {\bibfnamefont {G.}~\bibnamefont {White}},\ }\href@noop {} {\
  (\bibinfo {year} {2022}{\natexlab{b}})},\ \Eprint
  {http://arxiv.org/abs/2208.09789} {arXiv:2208.09789 [hep-ph]} \BibitemShut
  {NoStop}%
\bibitem [{\citenamefont {Durrer}\ and\ \citenamefont
  {Kusenko}(2022)}]{Durrer:2022cja}%
  \BibitemOpen
  \bibfield  {author} {\bibinfo {author} {\bibfnamefont {R.}~\bibnamefont
  {Durrer}}\ and\ \bibinfo {author} {\bibfnamefont {A.}~\bibnamefont
  {Kusenko}},\ }\href@noop {} {\  (\bibinfo {year} {2022})},\ \Eprint
  {http://arxiv.org/abs/2209.13313} {arXiv:2209.13313 [astro-ph.CO]}
  \BibitemShut {NoStop}%
\bibitem [{\citenamefont {Braaten}\ and\ \citenamefont
  {Pisarski}(1990)}]{Braaten:1990it}%
  \BibitemOpen
  \bibfield  {author} {\bibinfo {author} {\bibfnamefont {E.}~\bibnamefont
  {Braaten}}\ and\ \bibinfo {author} {\bibfnamefont {R.~D.}\ \bibnamefont
  {Pisarski}},\ }\href {\doibase 10.1103/PhysRevD.42.2156} {\bibfield
  {journal} {\bibinfo  {journal} {Phys. Rev. D}\ }\textbf {\bibinfo {volume}
  {42}},\ \bibinfo {pages} {2156} (\bibinfo {year} {1990})}\BibitemShut
  {NoStop}%
\bibitem [{\citenamefont {Bernal}\ \emph {et~al.}(2016)\citenamefont {Bernal},
  \citenamefont {Verde},\ and\ \citenamefont {Riess}}]{Bernal:2016gxb}%
  \BibitemOpen
  \bibfield  {author} {\bibinfo {author} {\bibfnamefont {J.~L.}\ \bibnamefont
  {Bernal}}, \bibinfo {author} {\bibfnamefont {L.}~\bibnamefont {Verde}}, \
  and\ \bibinfo {author} {\bibfnamefont {A.~G.}\ \bibnamefont {Riess}},\ }\href
  {\doibase 10.1088/1475-7516/2016/10/019} {\bibfield  {journal} {\bibinfo
  {journal} {JCAP}\ }\textbf {\bibinfo {volume} {10}},\ \bibinfo {pages} {019}
  (\bibinfo {year} {2016})},\ \Eprint {http://arxiv.org/abs/1607.05617}
  {arXiv:1607.05617 [astro-ph.CO]} \BibitemShut {NoStop}%
\bibitem [{\citenamefont {Gelmini}\ \emph {et~al.}(2021)\citenamefont
  {Gelmini}, \citenamefont {Kusenko},\ and\ \citenamefont
  {Takhistov}}]{Gelmini:2019deq}%
  \BibitemOpen
  \bibfield  {author} {\bibinfo {author} {\bibfnamefont {G.~B.}\ \bibnamefont
  {Gelmini}}, \bibinfo {author} {\bibfnamefont {A.}~\bibnamefont {Kusenko}}, \
  and\ \bibinfo {author} {\bibfnamefont {V.}~\bibnamefont {Takhistov}},\ }\href
  {\doibase 10.1088/1475-7516/2021/06/002} {\bibfield  {journal} {\bibinfo
  {journal} {JCAP}\ }\textbf {\bibinfo {volume} {06}},\ \bibinfo {pages} {002}
  (\bibinfo {year} {2021})},\ \Eprint {http://arxiv.org/abs/1906.10136}
  {arXiv:1906.10136 [astro-ph.CO]} \BibitemShut {NoStop}%
\bibitem [{\citenamefont {Anchordoqui}\ and\ \citenamefont
  {Perez~Bergliaffa}(2019)}]{Anchordoqui:2019yzc}%
  \BibitemOpen
  \bibfield  {author} {\bibinfo {author} {\bibfnamefont {L.~A.}\ \bibnamefont
  {Anchordoqui}}\ and\ \bibinfo {author} {\bibfnamefont {S.~E.}\ \bibnamefont
  {Perez~Bergliaffa}},\ }\href {\doibase 10.1103/PhysRevD.100.123525}
  {\bibfield  {journal} {\bibinfo  {journal} {Phys. Rev. D}\ }\textbf {\bibinfo
  {volume} {100}},\ \bibinfo {pages} {123525} (\bibinfo {year} {2019})},\
  \Eprint {http://arxiv.org/abs/1910.05860} {arXiv:1910.05860 [astro-ph.CO]}
  \BibitemShut {NoStop}%
\bibitem [{\citenamefont {Vattis}\ \emph {et~al.}(2019)\citenamefont {Vattis},
  \citenamefont {Koushiappas},\ and\ \citenamefont {Loeb}}]{Vattis:2019efj}%
  \BibitemOpen
  \bibfield  {author} {\bibinfo {author} {\bibfnamefont {K.}~\bibnamefont
  {Vattis}}, \bibinfo {author} {\bibfnamefont {S.~M.}\ \bibnamefont
  {Koushiappas}}, \ and\ \bibinfo {author} {\bibfnamefont {A.}~\bibnamefont
  {Loeb}},\ }\href {\doibase 10.1103/PhysRevD.99.121302} {\bibfield  {journal}
  {\bibinfo  {journal} {Phys. Rev. D}\ }\textbf {\bibinfo {volume} {99}},\
  \bibinfo {pages} {121302} (\bibinfo {year} {2019})},\ \Eprint
  {http://arxiv.org/abs/1903.06220} {arXiv:1903.06220 [astro-ph.CO]}
  \BibitemShut {NoStop}%
\bibitem [{\citenamefont {Escudero}\ and\ \citenamefont
  {Witte}(2020)}]{Escudero:2019gvw}%
  \BibitemOpen
  \bibfield  {author} {\bibinfo {author} {\bibfnamefont {M.}~\bibnamefont
  {Escudero}}\ and\ \bibinfo {author} {\bibfnamefont {S.~J.}\ \bibnamefont
  {Witte}},\ }\href {\doibase 10.1140/epjc/s10052-020-7854-5} {\bibfield
  {journal} {\bibinfo  {journal} {Eur. Phys. J. C}\ }\textbf {\bibinfo {volume}
  {80}},\ \bibinfo {pages} {294} (\bibinfo {year} {2020})},\ \Eprint
  {http://arxiv.org/abs/1909.04044} {arXiv:1909.04044 [astro-ph.CO]}
  \BibitemShut {NoStop}%
\bibitem [{\citenamefont {Gelmini}\ \emph {et~al.}(2020)\citenamefont
  {Gelmini}, \citenamefont {Kawasaki}, \citenamefont {Kusenko}, \citenamefont
  {Murai},\ and\ \citenamefont {Takhistov}}]{Gelmini:2020ekg}%
  \BibitemOpen
  \bibfield  {author} {\bibinfo {author} {\bibfnamefont {G.~B.}\ \bibnamefont
  {Gelmini}}, \bibinfo {author} {\bibfnamefont {M.}~\bibnamefont {Kawasaki}},
  \bibinfo {author} {\bibfnamefont {A.}~\bibnamefont {Kusenko}}, \bibinfo
  {author} {\bibfnamefont {K.}~\bibnamefont {Murai}}, \ and\ \bibinfo {author}
  {\bibfnamefont {V.}~\bibnamefont {Takhistov}},\ }\href {\doibase
  10.1088/1475-7516/2020/09/051} {\bibfield  {journal} {\bibinfo  {journal}
  {JCAP}\ }\textbf {\bibinfo {volume} {09}},\ \bibinfo {pages} {051} (\bibinfo
  {year} {2020})},\ \Eprint {http://arxiv.org/abs/2005.06721} {arXiv:2005.06721
  [hep-ph]} \BibitemShut {NoStop}%
\bibitem [{\citenamefont {Vagnozzi}(2020)}]{Vagnozzi:2019ezj}%
  \BibitemOpen
  \bibfield  {author} {\bibinfo {author} {\bibfnamefont {S.}~\bibnamefont
  {Vagnozzi}},\ }\href {\doibase 10.1103/PhysRevD.102.023518} {\bibfield
  {journal} {\bibinfo  {journal} {Phys. Rev. D}\ }\textbf {\bibinfo {volume}
  {102}},\ \bibinfo {pages} {023518} (\bibinfo {year} {2020})},\ \Eprint
  {http://arxiv.org/abs/1907.07569} {arXiv:1907.07569 [astro-ph.CO]}
  \BibitemShut {NoStop}%
\bibitem [{\citenamefont {Wong}\ \emph {et~al.}(2020)\citenamefont {Wong} \emph
  {et~al.}}]{Wong:2019kwg}%
  \BibitemOpen
  \bibfield  {author} {\bibinfo {author} {\bibfnamefont {K.~C.}\ \bibnamefont
  {Wong}} \emph {et~al.},\ }\href {\doibase 10.1093/mnras/stz3094} {\bibfield
  {journal} {\bibinfo  {journal} {Mon. Not. Roy. Astron. Soc.}\ }\textbf
  {\bibinfo {volume} {498}},\ \bibinfo {pages} {1420} (\bibinfo {year}
  {2020})},\ \Eprint {http://arxiv.org/abs/1907.04869} {arXiv:1907.04869
  [astro-ph.CO]} \BibitemShut {NoStop}%
\bibitem [{\citenamefont {Perivolaropoulos}\ and\ \citenamefont
  {Skara}(2022)}]{Perivolaropoulos:2021jda}%
  \BibitemOpen
  \bibfield  {author} {\bibinfo {author} {\bibfnamefont {L.}~\bibnamefont
  {Perivolaropoulos}}\ and\ \bibinfo {author} {\bibfnamefont {F.}~\bibnamefont
  {Skara}},\ }\href {\doibase 10.1016/j.newar.2022.101659} {\bibfield
  {journal} {\bibinfo  {journal} {New Astron. Rev.}\ }\textbf {\bibinfo
  {volume} {95}} (\bibinfo {year} {2022}),\ 10.1016/j.newar.2022.101659},\
  \Eprint {http://arxiv.org/abs/2105.05208} {arXiv:2105.05208 [astro-ph.CO]}
  \BibitemShut {NoStop}%
\end{thebibliography}%


\end{document}